# GALAXY DYNAMICS IN CLUSTERS

**Carlos S. Frenk**[1]

*Department of Physics, University of Durham*
*South Road, Durham DH1 3LE England*

**August E. Evrard**[1]

*Department of Physics, University of Michigan*
*Ann Arbor, MI 48109-1120 USA*

**Simon D.M. White**[1]

*Max-Planck-Institut für Astrophysik*
*Karl-Schwarzschild-Strasse 1, D-8046 Garching bei München, Germany*

**F J Summers**

*Department of Astrophysical Sciences*
*Princeton University, Princeton, NJ 08544 USA*

## ABSTRACT

We use high resolution simulations to study the formation and distribution of galaxies within a cluster which forms hierarchically. We follow both dark matter and a gas component which is subject to thermal pressure, shocks, and radiative cooling. Galaxy formation is identified with the dissipative collapse of the gas into cold, compact knots. We examine two extreme numerical representations of these galaxies during subsequent cluster evolution — one purely gaseous and the other purely stellar. The results are quite sensitive to this choice. Gas-galaxies merge efficiently with a dominant central object which grows to contain more than half of the galactic mass within the cluster. Star-galaxies merge less frequently and produce a mass distribution for cluster members which is quite similar in shape to that for non-cluster galaxies. Thus, simulations in which galaxies remain gaseous appear to suffer from an "overmerging" problem, but this problem is much less severe if the gas is allowed to turn into stars.

We compare the kinematics of the galaxy population in these two representations to the kinematics of dark halos and of the underlying dark matter distribution. Galaxies in the stellar representation are positively biased (*i.e.*, over-represented in the cluster) both by number and by mass fraction. Both representations predict the galaxies to be more centrally concentrated than the dark matter, whereas the dark halo population is more extended. A modest velocity bias also exists in both representations, with the largest effect, $\sigma_{gal}/\sigma_{DM} \simeq 0.7$, found for the more massive star-galaxies. Phase

[1] *Institute for Theoretical Physics, University of California, Santa Barbara, Ca 93106-4030*

diagrams show that the galaxy population has a substantial net inflow in the gas representation, while in the stellar case it is roughly in hydrostatic equilibrium. Virial mass estimators can underestimate the true cluster mass by up to a factor of 5 because of these various bias effects. The discrepancy is largest if only the most massive galaxies are used, reflecting significant mass segregation. A binding energy analysis suggests that this segregation is primarily a result of dynamical friction. We discuss the relevance of these results both to real clusters and to the general problem of simulating the formation and clustering of galaxies. The incorporation of a realistic star formation algorithm within future simulations will be the key to further progress.

## 1. Introduction

Galaxy clusters play a central role in cosmological studies. As the most massive nonlinear structures in the present Universe, they have been used to estimate the mean cosmic mass density and to constrain the nature of the dark matter. As young objects whose dynamical timescale is a large fraction of the age of the Universe, they have been used to probe the initial conditions for structure formation. Their evolution and present dynamical state are of considerable interest for studies both of large-scale structure and of the formation and evolution of galaxies.

Clusters are multicomponent systems in which dark matter, hot gas and galaxies evolve in a tightly coupled way. Their study is perhaps best approached through direct numerical simulation, but such simulations must include the proper cosmological context for cluster formation, as well as an appropriate representation of the three principal constituents and their interactions. In this paper we use simulations to explore the physical processes which establish the multicomponent nature of clusters, the mechanisms which determine the final distributions of the different components, and the extent to which nonlinear dynamical effects may prejudice the use of clusters as cosmological tools.

Simulations of cluster formation have gradually increased in complexity. Early N-body studies, beginning with Peebles (1970) and White (1976), concentrated on the collapse and relaxation of the dark matter component and on possible segregation effects acting on the most massive galaxies. Later work considered how the structure of clusters is related to the cosmological context in which they form (Quinn, Salmon & Zurek 1986; West, Dekel & Oemler 1987; Evrard 1987; White *et al.* 1988; Efstathiou *et al.* 1988; Frenk *et al.* 1990). Overdense regions separate from the general expansion, and the subsequent collapse destroys their clumpy initial structure to produce smooth, centrally concentrated configurations which are close to virial equilibrium. The final density profile depends on the initial spectrum of density fluctuations and on the mean cosmic density (Crone, Evrard, & Richstone 1994). Incomplete relaxation at intermediate times can result in substantial apparent substructure which may be used as an observational estimator of the cosmic density parameter, $\Omega$ (Richstone, Loeb & Turner 1992; Kauffmann & White 1993; Lacey & Cole 1993; Evrard *et al.* 1993).

The next level of complexity in cluster simulations was achieved through the inclusion of a collisional component to represent the intracluster gas. In the first such models hydrodynamic processes were included rather crudely by allowing inelastic collisions between gas particles (Carlberg 1988; Carlberg & Couchman 1989). Since large spatial and temporal variations in density occur during cluster evolution, the "smooth particle hydrodynamics" technique proves well suited to this problem. Simulations by Evrard (1990) and Thomas & Carlberg (1992) using $P^3M$/SPH codes, and simulations by Tsai, Bertschinger & Katz (1993) and by Navarro, Frenk & White (1995) using tree/SPH codes, showed that the collapse and shock heating of a nonradiative gas leads to an approximately isothermal equilibrium with a density distribution paralleling that of the dark matter. The X-ray emission produced by these models resembles that seen in real clusters, but the predicted distributions of cluster properties do not seem to conform to observation. For example, the predicted correlation between X-ray luminosity and gas temperature is too flat and evolution is in the opposite sense to that observed. The observed behaviour is much closer to that predicted for a nonradiative gas with fixed central entropy (Kaiser 1991; Evrard & Henry 1991), suggesting that nongravitational processes, such as feedback from galactic winds (White 1991; Metzler & Evrard 1994), have played a significant role in the evolution of the intracluster medium. Of course, radiative cooling is observed to play an important role in the inner regions of many clusters (*e.g.*, Fabian, Nulsen & Canizares 1994).

Neither N-body nor nonradiative SPH simulations treat processes related to the presence of galaxies in clusters. Although visible stars represent a negligible fraction of the total cluster mass (and only a small fraction of the directly observed mass, *e.g.*, White *et al.* 1993), it is evident that such processes should be included in any realistic cluster model. Some attempts have been made to study the effects of biased galaxy formation, dynamical friction, galaxy mergers and metal enrichment, either by inserting heavy "galaxy particles" by hand into the initial conditions (Evrard 1987; West & Richstone 1988) or by identifying their initial locations with high peaks of the linear overdensity field (White *et al.* 1986; Metzler & Evrard 1994). These calculations provide useful dynamical insights but they sidestep many issues concerning how galaxy and cluster formation are coupled. To address these questions, it is necessary to include additional processes such as gas cooling in a more explicit way. This is most directly done by carrying out much larger simulations which have sufficient dynamic range to follow both the collapse of the cluster material and the dissipative and collisional processes which regulate the formation and evolution of galaxies. Such simulations were carried out using the "sticky particle" treatment of gas dynamics by Carlberg (1988). It is now possible to attack the same problem with the much better treatment of hydrodynamics allowed by SPH. Such an investigation is the subject of this paper.

So far only a few cosmological N-body/SPH simulations have been published which include the effects of radiative cooling. The first was used by Carlberg, Couchman & Thomas (1990) to establish the concept of velocity bias. Katz, Hernquist & Weinberg (1992) explored galaxy formation in their study of evolution within a randomly chosen cubic region, 22 Mpc$^2$ on a side, in a universe dominated by cold dark matter (CDM). Working with $65,536$ particles and a gas particle mass of $1.2 \times 10^9 \mathrm{M}_\odot$, they showed that cooling of high density gas within dark matter halos led to the formation of tightly bound gas clumps which they identified with galaxies. The largest galaxy group in this simulation had 13 members with positions and velocities biased in such a way that a standard virial analysis underestimated the true group mass by a factor of about three.

Evrard, Summers & Davis (1994) used $524,288$ particles and a gas particle mass of $1.1 \times 10^8$ to simulate the evolution of a 16 Mpc cube within a CDM universe, and chose their box to enclose the expected formation site of a poor group. At their final time (which corresponded to $z = 1$), they found 26 "galaxy-like objects" in the central cluster. Again a standard virial analysis substantially underestimated the mass. Together with the simulation of Katz *et al* (1992), a lower resolution simulation by Katz and White (1994), and the "sticky particle" simulations of Carlberg & Couchman (1989), this work illustrates how cooling results in the formation of dense gas clumps which can survive the disruption of their halos within clusters. Thus, it supports the original conjecture of White & Rees (1978) that dissipative effects within hierarchical clustering theories can explain the existence of virialised clusters containing many distinct galaxies.

---

[2]We write Hubble's constant as $H_0 = 100\mathrm{h}^{-1}$ Mpc and, unless otherwise stated, we take $h = 0.5$.

An alternative approach to simulating the dynamics of gas in a cosmological context is being pursued by several groups (*e.g.*, Cen & Ostriker 1992a,b; Roettinger, Burns & Loken 1993; Kang *et al.* 1994; Bryan *et al.* 1994). This work uses finite difference techniques to follow the dynamics of a fluid which can undergo radiative heating and cooling and can form stars. Such grid-based methods are particularly useful for studying the large-scale gas distribution and for following dynamical effects in the neighborhood of shocks. However, the schemes used so far have had insufficient resolution to follow the details of cluster formation in a proper cosmological context or to simulate the formation and clustering of galaxies.

In this paper, we discuss results from new P3MSPH simulations which have a number of features in common with the model of Evrard *et al.* (1994). Like these authors, we follow the evolution of an initially overdense region, although our volume is larger and our initial conditions are generated in a different way. We concentrate on the dynamics of the galaxy population within the cluster, and we investigate how its apparent kinematic state depends on the way in which it is modelled. In particular, we compare results for two extreme cases, one where no star formation occurs in the cold dense clumps which are taken to represent galaxies, and the other where these clumps are turned into stars at an epoch well before cluster collapse. The differences between these two treatments turn out to be large. In examining these two extreme possibilities, we are motivated by the need to understand the behavior of the simulation techniques in relatively simple physical situations, as a prerequisite for including more complex phenomena such as star formation and the associated feedback processes. It turns out that even our simplified treatment provides some useful insights into the dynamical evolution of clusters.

The rest of this paper is organized as follows. In §2 we describe our simulation techniques and our procedure for generating initial conditions. In §3 we present an overview of the dynamical evolution of the cluster, including images to illustrate its global properties. In §4 we discuss the orbits and merger rates of the galaxies that form in our simulations. In §5 we analyse the evolution of the binding energy of galaxies and cluster dark matter and use this as a tool to understand the physical origin of various biases present in the cluster populations. In §6 we quantify the abundance of galaxies in the simulation and the efficiency of galaxy formation inside and outside the cluster. In §7 we investigate the structure, dynamics and equilibrium of the cluster galaxy population, the accuracy of virial mass estimates, and the implications for estimates of the mean cosmic density. Our paper concludes with a discussion of the main results in §8.

## 2. The Simulations

We use the P3MSPH code described by Evrard (1988; see also Efstathiou *et al.* 1985 and Summers, 1993) to model the evolution of 524288 particles. Half of them interact only through gravity and represent the dark matter and half of them experience both gravity and hydrodynamic forces and represent the gas. The gas is able to undergo adiabatic compression, shocks (as a result of an artificial viscosity term included in the hydrodynamic equations) and can cool radiatively, at a rate which depends on density and temperature according to the cooling function appropriate to an optically thin plasma of primordial composition in collisional ionization equilibrium. (The cooling function includes the effects of thermal bremsstrahlung, radiative recombination, dielectronic recombination and line emission, but not of heating by photoionization. This may be important on scales smaller than those we can resolve in our simulations (Efstathiou 1992)). The gas dynamic simulations described below were carried out using the Cray YMP8 at the San Diego Supercomputer Center and took roughly 150 hours of CPU time each.

### 2.1. Initial Conditions and Numerical Parameters

Our procedure for laying down initial conditions, designed to extend the dynamic range accessible to a single cosmological calculation, imposes a cluster perturbation in two basic steps. First, we identify a suitable cluster in an N-body simulation of a large region containing many clusters, in this case one of the simulations described by Frenk *et al.* (1990). These followed the evolution of 262144 dark matter particles in a box 360 Mpc on a side, assuming standard CDM initial conditions ($\Omega = 1$, $h = 0.5$ and normalization at $8h^{-1}$ Mpc, $\sigma_8 = 0.59$). The chosen cluster had a one-dimensional velocity dispersion of 800 km s$^{-1}$ and a "turnaround radius" of $\sim 12$ Mpc. The displacements due to large-scale waves in the initial conditions were replicated in a smaller box, of comoving size 45 Mpc, centered on the cluster and loaded with 262144 dark matter particles on a $128^3$ grid. The second step perturbs this particle distribution with additional waves chosen to represent the CDM power spectrum between the original resolution limit and the Nyquist frequency of the new particle grid. The resulting displacement distribution is then apodized with a cosine bell in order to avoid discontinuities at the periodic boundaries of the box. (This last step modifies the initial displacements of particles within 1.5 comoving Mpc of the box boundary). The resulting initial conditions have a mass resolution about 20 times better than the original simulation. Initial velocities were assigned according to the Zel'dovich approximation in the manner described by Efstathiou *et al.* (1985). Evolving this dark matter configuration with a P$^3$M code produced a cluster similar to the original one but with additional substructure due to the extra high frequency waves.

Ten percent of the mass of each dark mass particle was then removed and placed in a superposed gas particle with identical position and velocity, and a temperature of $10^4$K. This leads to a simulation with $\Omega_b = 0.1$. In our first simulation, a fixed comoving gravitational softening parameter $\epsilon = 35$ kpc was used with a Plummer potential, $\phi(r) \sim (r^2 + \epsilon^2)^{-1/2}$. Within P3MSPH, the SPH smoothing parameter, $h$, is never allowed to fall below $0.5\epsilon$ while its upper limit is set by the size of the P$^3$M chaining mesh. This results in an effective spatial resolution in the densest regions of $\simeq 2\epsilon$. The simulation was evolved from redshift $z_i = 12.4$ to the present over 700 equally spaced timesteps.

## 2.2. Resolution Effects

This first experiment failed to produce an appreciable number of galaxies. By redshift $z = 0.18$, when the simulation was stopped, gas had been able to cool onto only four distinct objects containing 32 or more particles and with density contrast $\sim 10^6$. We interpret this deficiency of galactic size objects as a result of inadequate mass resolution. Each gas particle in this model has a mass of $2.4 \times 10^9$ M$_\odot$, a factor 24 larger than the corresponding mass in the experiment of Evrard *et al.* (1994). An $L_*$ galaxy would be resolved by a clump of only a few tens of particles, right at the resolution limit of the experiment. Real clusters of this mass contain roughly 50 $L_*$ galaxies. Since analytic work by Kauffmann *et al.* (1993) shows that the kind of CDM model we are studying should produce approximately the right number of such galaxies, the problem would appear to be that the gas associated with small dark matter clumps is unable to cool as efficiently as expected.

To check this hypothesis, we ran a second simulation using the same initial displacements and velocities, but scaling the box size down by a factor of 2 (to 22.5 Mpc) and rescaling all other physical quantities appropriately. The mass of an individual gas particle then becomes $3 \times 10^8$ M$_\odot$, the gravitational softening becomes $\epsilon = 17.5$ kpc, and an $L_*$ galaxy is represented by a knot of about 350 gas particles. In the absence of radiative cooling the evolution of the two simulations would be identical. However, the rescaling leads to a reduction of a factor of 4 in the temperature of all collapsed objects and this, in turn, leads to a substantial reduction in the cooling time of the gas in the first objects which form. As a result this second simulation (hereafter referred to as the "gas" simulation) produced a few hundred dense, cold objects of galactic mass (the "globs" of Evrard *et al.* 1994). Notice that this test proves that the lack of galaxies in the first model is a result of inefficient cooling of objects rather than of any inability to represent their formation in our SPH scheme. Examination of Figure 1 below suggests that the difficulty reflects the fact that objects made up of only a few tens of particles tend to be significantly less concentrated in the gas component than in the dark matter whereas studies of individual objects carried out with much higher resolution suggest that the cooled gas should, if anything, be more concentrated.

While this rescaling provides a clean test of the source of the problem in our original model and leads to a final cluster with a large enough population of galaxies for its properties to be studied in detail, it is difficult to place the new model in a plausible cosmological context. After rescaling, the linear power spectrum at the final time corresponds roughly to that of a CDM universe with $\sigma_8 \sim 0.3$. As a result, the formation history of the cluster is rather atypical of those expected for clusters with a velocity dispersion of 400 km/s in a more standard CDM universe (*e.g.*, $\sigma_8 = 0.6$). In particular, the precursor structures to the cluster (including the galaxies) form and merge significantly later than in more standard models. In what follows we stick to the rescaled variables when analysing this simulation and the star-galaxy simulation derived from it (see below). However, it is worth noting that another valid interpretation is to retain the original scaling and to argue that the second model compensates for the suppression of cooling by resolution effects in an *ad hoc* way by calculating a cooling time for each particle using a temperature which is one quarter of the true value. In this interpretation all sizes and velocities quoted in the rest of the paper should be doubled, all temperatures should be quadrupled, and all masses should be multiplied by 8. Densities, times and redshifts are unaffected.

## 2.3. Dynamical Treatment of the Galaxy Population

The main aim of this paper is to investigate whether gas dynamical simulations can produce realistic cluster models. Real galaxies are, of course, made predominantly of stars rather than of cold gas. We therefore designed an experiment to test the dynamical consequences of neglecting star formation. At a redshift of 0.7, well before the collapse of the main cluster but well after a significant amount of gas has condensed into cool clumps, we identified candidate galaxies by using a friends-of-friends grouping algorithm on the gas particles (Davis *et al.* 1985). We adopted a linking length which was only 1.7% of the mean interparticle separation, and so identified objects at very high density contrast ($\sim 10^5$). All the gas particles in groups with 32 or more members were instantaneously converted into collisionless "star" particles. All the remaining gas was removed and the mass associated with it was distributed uniformly among nearby dark matter particles. The resulting distribution of "stars" and dark matter was evolved from $z = 0.7$ to $z = 0$ with a collisionless P$^3$M code. We will refer to this calculation as the "star" simulation and to collapsed objects containing stars as "S-gals" to distinguish them from the purely gaseous objects in the original simulation which we refer to as "G-gals."

In the analysis of the later evolution of both these simulations, we define galaxies by using the group finder on the appropriate particle distribution, with a linking length of $(1+z)\%$ of the mean interparticle separation. This procedure picks out regions which lie above a fixed *physical* density threshold of roughly $0.5 \mathrm{cm}^{-3}$. The procedure is identical to that used by Evrard *et al.* (1994).

## 3. Overview of the Simulations

Figure 1 shows the distribution of dark matter (left panels) and gas (right panels) in a slice with sides one half and thickness one–tenth that of the total volume in the "gas" simulation. Three different epochs are shown, corresponding to redshifts $z = 2, 0.7$ and 0. Both the dark matter and the gas fall coherently onto the central density enhancement. The flow is highly inhomogenous and a complex filamentary pattern, converging at the centre, is clearly visible. In spite of these high contrast features, the flow is never highly anisotropic: as we shall see below, the particles that end up in the cluster come from a roughly spherical initial region. At $z = 0.7$, several large subcondensations have formed. These all merge by $z = 0$, but even then the cluster does not yet appear fully relaxed. As the large subclumps come together, their dark matter interpenetrates, but their gas is shocked. Energy is transferred during the collision, both from the dark matter to the hot gas (Navarro & White 1994; Pearce, Thomas & Couchman 1994) and from the cold to hot phases of the gas itself due to viscous braking of infalling G-gals. As a result, the hot gas ends up being slightly more extended than the dark matter. Because of its (isotropic) pressure support, the gas also has a rounder configuration. Notice that the dense clumps of cold gas, while present, are not readily visible in Fig. 1 due to their extremely small volume filling factor.

## 3.1. The Assembly of the Cluster Material

Our working definition of the cluster is the material within a sphere of radius 2 Mpc centered on the largest dark matter clump at $z = 0$. The mean mass density contrast within this sphere is $\sim 110$. Although this is slightly larger than the fiducial cluster radius (within which $\delta\rho/\rho = 180$) we shall see in §6 below that some material which has passed through the cluster center lies beyond 2 Mpc. Clusters have no well defined edge in a cosmological setting, so any measure of size is, to some degree, arbitrary.

Figure 2 illustrates the time evolution of the material that ends up in the body of the cluster. Distributions of dark matter, hot gas ($T > 3 \times 10^6$K), cool gas ($T < 3 \times 10^6$K), G-gals, and S-gals are plotted. Time runs from left to right and the six epochs shown correspond to redshifts $z = 2, 1, 0.7, 0.3, 0.1$ and 0 respectively. The cluster is assembled in a very lumpy fashion. The large subcondensations seen at early times in Figure 1 fall together and are disrupted between $z = 0.3$ and $z = 0.1$, producing a diffuse dark matter background containing smaller, high contrast lumps. By $z = 0.1$, most of the large lumps have disappeared, but a few smaller dark matter halos still survive. By $z = 0$, the dark matter distribution is smooth and centrally concentrated, and most of the remaining galactic halos lie in the periphery of the cluster.

At early times, the gas is (by assumption) cold, and so it clusters with the dark matter. As soon as dark matter halos form which are sufficiently large to be resolved by the SPH technique, their associated gas falls to halo center where it settles into a cold, centrifugally supported disk (Evrard *et al.* 1994). During such collapses some of the gas is shock-heated to form smooth hot coronae. The largest of these (those associated with dark matter clumps with virial temperatures above $3 \times 10^6$K) may be seen in Figure 2b. As the cluster grows, an increasing fraction of the cold gas collects in small nonlinear clumps, but much remains diffuse or in hot coronae and these components merge until, by the present day, most of the gas in the cluster is in a hot rarefied atmosphere near hydrostatic equilibrium and with no appreciable substructure.

The first G-gals form before $z = 2$ and their abundance grows rapidly so that by $z = 0.7$, when large precluster condensations are collapsing, there is already a sizeable population of them. It is at this time that we turn G-gals into S-gals in the "star run." The subsequent evolution of the galaxy populations in the two runs is rather different. The G-gals experience viscous interactions and their collisions are sticky. As a result, they tend to merge into a single massive object at the center of the cluster which eventually contains almost half the cold gas in the cluster. Occasionally G-gals are removed from their halos and disrupted as they fall through the hot gas. An example of this process can be seen near the top left-hand corner of the $z = 0$ panel of Figure 2c. High speed collisions between G-gals in the central regions of the cluster also lead to the disruption of a significant number of systems – more than 30% of the mass in cluster G-gals is lost to the diffuse medium in this way between $z = 0.3$ and the end of the simulation. In Sections 4 to 6 below we compare how the G- and S-gal populations evolve, and we compare both to the population of dark halos.

Further details of the evolution of the hot gas are illustrated in Figure 3. The first three columns show contour diagrams of the projected dark matter distribution, gas density and gas temperature within a physical region of size 3.3 Mpc. The last column shows contour maps of X–ray luminosity in the ROSAT passband $(0.5 - 2.4$ keV) obtained from the density and temperature of the gas in the manner described by Evrard (1990). These X-ray images are what a hypothetical satellite would see if situated a fixed distance of 180 Mpc (corresponding to an effective source—observer redshift of 0.03) from the cluster at all epochs. The epochs shown correspond to redshifts of 0.7, 0.3, 0.1 and 0.03. Qualitatively, the evolution of the gas distribution resembles that of the dark matter, but shocks lead to transient features in the gas which are visible in the density and temperature plots. Note how the gas is compressed and heated at the interface of the two large merging subunits at $z = 0.3$, producing a stream of hot, diffuse gas which squirts in the direction perpendicular to the axis of the collision. At $z = 0$ the gas near the center remains quite inhomogeneous as it cools and sloshes around in the varying gravitational potential. The rapid decline in the population of small dark matter halos is easily seen in the projected mass distribution. Note that by $z = 0.03$ very few such halos remain, most of those in the outskirts of the cluster. The largest G-gals show up in the density and temperature plots as dense condensations of low entropy gas. Small galaxies are not resolved in this projected image.

The X-ray images of Figure 3 clearly illustrate the dynamically complex nature of cluster formation. At $z = 0.7$, the cluster is broken up into several distinct subunits with a wide range of sizes. Cooling is important in the smaller knots which appear more centrally concentrated than the two dominant components. A few tightly bound subunits of high surface brightness survive at $z = 0.3$, when the central region of the cluster shows a clear double structure. By $z = 0.1$, the X-ray isophotes are still irregular and have the "boxy" appearance characteristic of a major ongoing merger. A bright subclump infalls into the central regions from the lower right. Note the factor of $\sim 2$ drop in the projected temperature map across the interface between the main body of the cluster and the subclump. By $z = 0.03$, this subclump has merged with the main cluster and the X-ray appearance of the cluster is fairly regular, with outer isophotes somewhat elongated along the principal collision axis. At $z = 0.03$, the total X-ray luminosity (in the ROSAT band) emitted by the region shown is $6.6 \times 10^{43}$ erg s$^{-1}$ ($4.9 \times 10^{44}$ erg s$^{-1}$ if the simulation is rescaled to the original 45 Mpc box size).

## 4. Galactic Orbits and Merger History

The differences in the abundance and final distributions of G-gals and S-gals apparent in Figure 2 reflect the different dynamical behavior of these two types of objects. Since the masses, positions and velocities of G- and S-gals are identical at $z = 0.7$, differences in their interactions with each other and with the intracluster medium must give rise to their divergent evolution. The orbits of the 16 most massive G- and S-gals are plotted in Figure 4. To trace the orbits, the membership lists of groups at successive outputs are compared, and a group at the later time is identified with one at the earlier time if more than half of the latter's members are included in the former. The time interval between outputs is approximately $2 \times 10^8$ yr. The positions of the center of mass are plotted as dots in the figure; the cross marks the position of the centre of the cluster at $z = 0$. The number in each panel gives the mass rank of the object at $z = 0.7$. Gaps in the ranks arise from objects which lie outside the body of the cluster. Notice that several of the orbits plotted are identical at later times as a result of merging of the objects considered.

The orbits of the G-gals and S-gals are different, though the degree of difference depends on mass. Both types of object have predominantly radial orbits, but as they approach the central few hundred kiloparsecs, their trajectories diverge. G-gals experience a viscous interaction with the intracluster medium and dissipative collisions with each other. Most of them do not re-emerge from the cluster core after their first pericentric passage, and they often merge with the central gaseous object. The S-gals, on the other hand, do generally pass through the cluster core, and several of them survive several pericentric passages. Nevertheless, even the S-gals finish relatively few orbits; the crossing time of the present-day cluster at 2 Mpc, $t_c = \pi R^{3/2}(2GM)^{-1/2} \sim 6 \times 10^9$ yrs, is comparable to the time elapsed since $z = 0.7$. The overall impression from these plots is not one of an old, relaxed system of galaxies swarming around in a fixed cluster potential, but rather one of a young and dynamically evolving system.

A possible concern is that numerical effects might artificially enhance the viscous drag on G-gals. Order of magnitude estimates and explicit numerical tests show that viscous drag against the diffuse gas should not be important in the outer cluster. In Figure 5 we plot the time evolution of the distance between a G-gal or its corresponding S-gal and the current position of the most massive G-gal. The latter marks the density center of the gas run. Galaxies follow similar trajectories in the two runs until they first pass near the cluster center. Thus the braking of G-gals appears to take place in the inner few hundred kiloparsecs, where the gas density in the hot phase is high ($\sim 0.01 \text{cm}^{-3}$) and where other G-gals occupy a substantial fraction of the volume. In most cases the orbits diverge rapidly as they pass through the central region. In several instances (e.g., objects 11, 14 and 23), the G-gal is left behind near the center while the S-gal continues along an orbit which takes it back to a radius $\sim 1$ Mpc. In other cases the S-gal remains on an orbit a few hundred kiloparsecs from cluster center while its G-gal counterpart merges with the central galaxy. As a result the final dominant objects in the centers of the two runs are very different. Figure 6 displays their merger histories, generated by linking groups from one time to the previous one; a time reversal of the procedure used to generate Figure 4. At $z = 0.7$, there are 24 objects which ultimately merge to form the dominant G-gal. In contrast, the largest S-gal forms from a single merger between two nearly equal mass objects whose relative orbit decays gradually over $\sim 2.5$ billion years.

To illustrate the dramatic consequences of these effects for the final appearance of the cluster, Figure 7 shows an image of the central 2.25 Mpc region for each representation of the galaxies. In these plots the circle marking each galaxy has an area proportional to its mass. The largest G-gal dominates the "light" of the cluster and lies smack on the cluster center. The largest S-gal is not nearly so dominant and it lies about 200 kpc away from the projected cluster center. This figure clearly shows how the G-gals end up in a much more compact configuration than the S-gals.

To summarize, most G-gals which traverse the cluster core merge into the central object, whereas S-gals typically survive much longer and rarely merge. Thus previous SPH simulations which identified galaxies as cold, dense gas knots (*e.g.*, Carlberg *et al.* 1990; Katz *et al.* 1992; Evrard *et al.* 1994) are unlikely to provide an acceptable description of the masses or of the clustering of galaxies. Our transformation of G-gals into S-gals is clearly too *ad hoc* to be considered a realistic model for galaxy formation; a reliable derivation of the masses and clustering of galaxies will require a much more careful treatment of galaxy formation than has been attempted by us (or by anyone else) so far.

## 5. Binding Energy Analysis

Further clarification of the physical processes that induce differences between the distributions of G-gals, S-gals and dark matter halos can be obtained by tracking the evolution of their orbital binding energy. A variety of factors affect this energy. For example, in the "high peak model" galaxies are predicted to form more efficiently than average in protocluster regions and so may be born with systematically greater binding energy than random dark matter particles (Kaiser 1984, Davis *et al.* 1985, Bardeen *et al.* 1986). Later dynamical processes can impose additional biases. For example, dynamical friction causes massive galaxies to lose energy to the dark matter and to sink towards the cluster center, while galaxy merging can reduce the total number of galaxies and increase their luminosity. Differences present at early times may reasonably be ascribed to statistical "high peak" biases, whereas differences produced during and after cluster collapse must be of dynamical origin.

We calculate binding energies by convolving the mass distribution with a Plummer potential evaluated on a $128^3$ mesh with a softening value of 0.225 Mpc. The potential at the position of each particle is then obtained by cloud-in-cell interpolation, and its binding energy is estimated by adding its specific kinetic energy. This procedure smooths over the contributions to the binding energy from very small-scale structures, allowing us to examine biases in the large–scale cluster potential.

Distributions of specific binding energy for "cluster particles" (*i.e.*, particles which lie within 2 Mpc of the cluster center at $z = 0$) are plotted for several different epochs in Figure 8. (The binding energy is given in arbitrary units and is negative for bound particles.) The left-hand panel compares the binding energy of the dark matter to that of the S-gals, while the right-hand panel compares the binding energies of the G-gals and S-gals. (Only G- and S-gals with $N \geq 32$ were considered; note that the G-gal and S-gal histograms are mass-weighted.) The collapse of the cluster is manifest in the dark matter plot as the rapid decrease in binding energy between $z = 0.7$ and $z = 0.3$. After this time, the potential well of the cluster continues to deepen but at a slower rate. The evolution of the binding energies of the S-gal and G-gal populations are qualitatively similar to this. However, already at $z = 0.7$, the binding energy distributions of these populations are biased towards more negative values than that of the dark matter. Since at this epoch the cluster has not yet collapsed, this bias may be interpreted as a property of the "initial conditions". Note that this is not a negligible effect. At the median of the distributions, the difference in binding energies between dark matter and S-gals at $z = 0.7$ is about 12%, compared to about 25% at $z = 0$.

By construction, the binding energy distributions of the G-gal and S-gal populations (right-hand panel of Figure 8) are almost identical at $z = 0.7$. (The small difference reflects the fact that there are a few S-gals which lie just outside 2 Mpc at $z = 0$ while their G-gal counterparts are just inside.) The collapse of the cluster at $z \simeq 0.3$ affects the two populations in a similar way, but by $z = 0$ the population of G-gals is considerably more tightly bound than the population of S-gals. This reflects the energy losses experienced by the G-gals through viscous interactions and leads to their more centrally concentrated distribution relative to the S-gals as discussed above in Section 4.

It is instructive to examine in some detail the origin of the differences between the binding energies of the dark matter and collapsed clumps. Particularly revealing are the trajectories, in binding energy space, of dark matter particles and individual clumps of similar initial binding energy. Figure 9 compares the binding energies at two different epochs of a randomly chosen subset of cluster dark matter particles. (The curves are logarithmically spaced isodensity contours and will be discussed below.) The top panel shows how the binding energies change between $z = 0.7$ and $z = 0.3$, the interval during which much of the cluster is assembled, but the evolution is relatively mild. The binding energies at these two epochs are well correlated, but with considerable scatter. Infalling material shows up as a cloud of points in the top right-hand corner of the diagram. The bottom panel of Figure 9 shows how the dark matter binding energy changes over the entire redshift range, $z = 0.7$ to $z = 0$. There are only minor differences between this and the top diagram, reflecting our earlier conclusion that the overall cluster potential evolves little between $z = 0.3$ and $z = 0$, even though major subcluster units are merging during this interval.

Figure 10 contrasts the evolution of the dark matter with that of the G-gals and S-gals. All cluster particles which are members of a "galaxy" at both the plotted redshifts as well as at $z = 0$ are shown, while the dark matter is represented by the isodensity contours from Figure 9. The galaxies appear as distinct clumps. Clumps corresponding to galaxies that merge during the interval considered line up at a fixed binding energy at the later redshift. Many galaxy clumps are stretched in the vertical direction. In the case of the S-gals, this reflects heating by tidal and 2-body effects between $z = 0.7$, when the cold gas was converted to stars, and the later epoch when the clumps are examined again. In the case of the G-gals, vertical stretching reflects a recent merger which has temporarily led to a substantial increase in the random motions of gas particles within the merging clumps.

At $z = 0.7$, the distributions of S-gals and G-gals in Figure 10 are slightly shifted to the left relative to the dark matter distribution, reflecting the "initial" bias seen in Figure 8. Between $z = 0.7$ and $z = 0.3$, the G- and S-gals behave roughly like the dark matter, but there are some interesting differences of detail. For example, the G-gals become more distended than the S-gals primarily because they undergo more mergers. (A notable exception is the most massive S-gal which has a large velocity dispersion both at 0.3 and at 0.0 and so a large vertical extent in the plots.) In addition, by $z = 0.3$ a number of mergers of G-gals are apparent which have no S-gal counterparts.

The lower panels of Figure 10, which compare binding energies at $z = 0.7$ and at $z = 0$, reveal quite dramatic differences among the three components. Material which had the same binding energy at $z = 0.7$ can end up with an entirely different binding energy at $z = 0$, depending on whether it is dark matter, an S-gal, or a G-gal. The segregation relative to the dark matter is much greater for the G-gals than for the S-gals, but even the S-gals show a marked tendency for the largest clumps to end up near the center of the potential well. Virtually all the G-gals which survive until $z = 0$ have merged into two dominant objects; the only other surviving objects were loosely bound initially and are currently falling in for the first time (see Figure 18 below). The S-gals are spread over a larger range of binding energy at $z = 0$ than the G-gals, but also show a marked deficit of moderately bound members compared to the dark matter. The distributions of G-gals, S-gals and dark matter are similar in the outer parts of the cluster where the ridge line of accreting objects is visible in all three components. These plots suggest that most of the segregation between massive clumps and dark matter occurs after $z = 0.3$, when much of the cluster material is already in place.

These diverging evolutionary trajectories presumably reflect the different interactions to which the three components are subject. S-gals experience only gravity, but their relatively high mass causes them to lose energy to the diffuse dark matter background. It seems appropriate to refer to this mechanism as dynamical friction even though the transfer of energy between the populations in a lumpy and rapidly changing environment does not conform to the standard picture in which a massive object gradually spirals to the center of a quasistatic potential well. The stronger biases apparent for the more massive S-gals (see Table 3) favor this interpretation, since dynamical friction should increase linearly with galaxy mass. The G-gals also experience dynamical friction, but their evolution is further influenced by the viscous effects which enhance the tendencies towards merging and gravitational settling. These processes combine with the bias present before collapse to produce the strong mass segregation seen in Figure 7 and an associated velocity bias which we discuss in Section 7 below. As we shall see, these biases cause a standard virial analysis to substantially underestimate the mass of the cluster.

## 6. Galaxy Formation Efficiency

In this section, we examine how the cluster environment affects the abundance of galaxies and their mass distribution. In hierarchical models, the abundance of objects of a given mass is determined by competition between the formation of new objects and the destruction of existing ones as the merging hierarchy builds up. Since the rates of these processes depend on the large-scale environment, one might expect the characteristics of galaxies inside and outside of clusters to differ.

At each epoch, we identify G-gals or S-gals through the procedure described in Section 2.3, *i.e.*, by using a friends-of-friends grouping algorithm with linking length $\eta = 0.01(1 + z)$. We also examine dark matter halos, but we identify them in a slightly different manner. Lacking dissipation, the dark matter does not achieve the high density contrasts seen in the baryons. At $z = 0$, for example, only the center of the main cluster is found with a linking parameter $\eta = 0.01$ for the dark matter, while there are nearly 200 G-gals found with 32 or more particles. We therefore identify dark halos with a redshift–independent linking length equal to 5% of the mean interparticle separation. This picks out objects at a fixed density contrast of roughly 8000. The distribution of dark matter halos at various redshifts is shown in Figure 2(f). This may be compared to the corresponding distributions of G-gals and S-gals, also plotted in Figure 2.

Note that the number of dark matter halos associated with the cluster is quite sensitive to the choice of linking parameter, but the number of G- and S-gals is not. The latter all have very high density contrast and are easily picked out by the group finder. Substructure in the dark matter, on the other hand, spans a wide range of density contrasts and so different populations are picked out by different values of $\eta$. For the simulation as a whole, our choice of $\eta = 0.05$ yields a mass fraction in dark halos with 32 or more particles which is similar to the fraction of the gas in the form of G-gals.

## 6.1. Number Density and Mass Fraction of Collapsed Objects

Figure 11 shows the number of objects as a function of redshift, both in the entire volume and in the cluster material. As discussed in Section 3.1, we define cluster objects to be those made up of particles which, at $z = 0$, are contained within a sphere of radius 2 Mpc centred on the largest dark matter condensation. Complementary information is provided by the plot in Figure 12 which shows the evolution of the cumulative mass fraction in collapsed objects. For cluster material, we use a Lagrangian definition of the mass fraction:

$$f(z) = \frac{\sum_{N_i \geq N_{min}} N_i(z)}{N_{DM}}, \qquad (1)$$

where $N_i(z)$ is the number of particles in objects identified at redshift $z$ from material tagged within 2 Mpc of the cluster at the present epoch. The normalization constant, $N_{DM}$, is the number of dark matter particles within the cluster at $z = 0$; for the gas run, $N_{DM} = 85018$, while for the star run, $N_{DM} = 83836$. Three different mass ranges are shown in each plot, corresponding to particle numbers, $N \geq 32$, 128 and 512.

The G-gal and S-gal populations are identical at $z = 0.7$, by construction. (The slightly different number of G- and S-gals at $z = 0.7$ in the cluster panel of Figure 11b is an edge effect. For the reasons discussed in Section 4, a few S-gals pass through the cluster center and end up on orbits stretching beyond 2 Mpc, whereas their G-gal counterparts lie within the 2 Mpc cutoff at $z = 0$.) After $z = 0.7$ new halos and G-gals can form by collapse and, for the latter, by gas cooling, but these processes are only important outside the body of the main cluster. In the volume as a whole the number of resolved G-gals grows by $\sim 20\%$ to a peak of 240 at $z = 0.3$ before declining again. This decline is due to the rapid merging of objects in the cluster, where the number of G-gals with 32 or more particles drops by a factor 3 between $z = 0.7$ and the present. As we have seen this merging produces a a large, central galaxy containing over 50% of the galactic mass of the cluster. The importance of mergers is also obvious as an increase in the mass fraction contained in the heaviest G-gals in the cluster (central panel of Figure 12b.) The drop in the cumulative mass fraction at the resolution limit ($\geq 32$ particles) indicates that $\sim 30\%$ of the G-gal material is lost as tidal or collisional debris, and that little new material is added to the cluster galaxies by cooling. By the present epoch, 12% of the total gas mass resides in G-gals.

Since no new low-mass S-gals can form after $z = 0.7$, their total numbers can only decline through merging and tidal disruption. The merging of S-gals is much less efficient than that of the G-gals. The most massive S-gals form exclusively in the cluster and the number of objects with more than 512 particles grows from one initially to a maximum of three before dropping to a final value of two. The cumulative mass fraction for all but the most massive S-gals also declines slowly over the course of the run. For minimally resolved objects, disruption effects cause a 20% loss within the entire volume and a $\sim 30\%$ loss for S-gals within the cluster. At the present epoch, the mass fraction in S-gals is 6%, half that in G-gals.

The dark matter halos exhibit a somewhat different behavior. Like the G-gals, the number of minimally resolved objects in the volume grows with time due to the continual collapse of small perturbations. The number in the most massive bin varies little until the major merger event which assembles the cluster at $z \simeq 0.2$. At this point, the number of large dark matter halos in the cluster declines from four to one (bottom of Figure 11b). The oscillation in the mass fraction associated with cluster dark matter halos (top panel of Figure 12b) is due to the relaxation of the main cluster after the violent merger event. It is clear that dark halos are not completely destroyed within the cluster. At $z = 0$, there are nearly thirty halos with 32 or more particles inside 2 Mpc, although, as we will see in Section 7, most of them reside quite far from the cluster center.

## 6.2. The mass function of collapsed objects

The present-day differential mass functions for G-gals and S-gals within the entire simulated volume are displayed in Figure 13. Except at the very massive end, they are quite similar. Both are moderately well fit by the Schechter function which is often used to describe the observed luminosity function of galaxies,

$$\Phi(L/L_*)dL/L_* = \Phi_* (L/L_*)^\alpha \exp{-(L/L_*)} dL/L_*. \qquad (2)$$

The slope at the low mass end in Figure 13, $\alpha \simeq -1.8$, is much steeper than the slope at faint end of the observed galaxy luminosity function, $\alpha \simeq -1$ (Loveday et al. 1992; but compare Marzke et al. 1994). This discrepancy is a well known feature of hierarchical clustering models of the kind assumed here. It has been noted before both in analytic work (White & Rees 1978; White & Frenk 1991; Cole 1991) and in numerical simulations (Evrard et al. 1994) and a number of plausible explanations have been proposed. These range from the possibility that the observed function may be biased low if low surface brightness galaxies are missed or if their total luminosities are underestimated in magnitude limited catalogues (e.g., Disney and Phillipps 1988, Bothun, Impey & Malin 1991; Ferguson & McGaugh 1995; Dalcanton, Spergel, & Summers 1995) to the suggestion that galaxies in small halos may form inefficiently as a result of strong feedback effects (Larson 1974; Dekel & Silk 1986; Lacey et al. 1993; Kauffmann et al. 1993; Cole et al. 1994). The mass function within our cluster is similar to that in the simulation as a whole (see fig. 15 below). There is some observational evidence that the faint end of the luminosity function in clusters may indeed be quite steep (Driver et al. 1994; Biviano et al. 1995; Bernstein 1995).

## 6.3. Environmental Biases

Since the efficiencies of galaxy formation and of merging differ in the cluster and in the simulation as a whole, the ratio of the number of collapsed objects to the amount of dark matter is a function of environment. Let us define a "number bias" as the ratio of the number of cluster objects of a given population (Figure 11b) to the number predicted based on statistics for the simulation as a whole,

$$N_{exp} = N_{vol} \frac{N_{DM}}{64^3}, \tag{3}$$

where $N_{vol}$ is the number of objects of a given type found in the entire volume, $N_{DM}$ is the cluster dark mass defined above, and $64^3$ is the number of dark matter particles in the simulation. The mass fraction bias is similarly defined as the ratio of the mass fractions shown in Figures 12b and 12a. The growth with time in the bias in the G-gal and S-gal populations is shown in Figure 14. Note that a maximal bias of $64^3/N_{DM} \sim 3.1$ would be achieved if objects of a given type were found *only* in the cluster.

We caution that the simulation as a whole is certainly not a fair representation of the "field" population. Radial gradients in the galaxy populations extend outward from the cluster center all the way to the simualtion boundary. In addition, the apodization required at the boundary affects fluctuations in roughly 20% of the volume. As a result we do not stress the overall magnitude of the biases, but rather focus on their qualitative behavior.

As may be seen in Figure 14, the S-gals retain a roughly constant bias in both a number- and a mass-weighted sense. This results from the fact that cluster S-gals are not significantly depleted by tidal stripping or mergers. Notice that the bias they inherit from the G-gals at $z = 0.7$ is mass-dependent, with the most massive objects found almost exclusively in the protocluster. On the other hand, although the G-gals begin with an identical distribution, both their number and their mass fraction biases decline significantly after the merger which assembles the bulk of the cluster. By number, the G-gals in *all* mass ranges are anti-biased at the final epoch, much of the change being caused by mergers within the cluster. The decline in their mass fraction bias results from mass loss due to viscous stripping inside the cluster and mass gain by cooling and accretion outside the cluster.

The resultant mass functions at $z = 0$ are displayed in Figure 15. The solid lines show the actual cluster mass functions while the dashed lines show the expectations derived from the entire volume by renormalizing by the factor $N_{DM}/64^3$. Not surprisingly, the dark matter halo population in the cluster is dominated by a single large object containing over 80% of the mass in halos. Although the heaviest G-gals are only found in the cluster, the number bias is positive only for objects with more than $\sim 550$ particles, corresponding to a baryon mass of $1.6 \times 10^{11} M_\odot$. The S-gals are the only population which remains positively biased in both number and mass weighted senses over all mass ranges resolved. Table 1 lists the masses of the top three objects of each type found in the cluster at $z = 0$. For the dark matter halos, the ratio of the first to second ranked objects is $M_1/M_2 = 34.1$. The G-gals are also top–heavy, but with a much smaller value of $M_1/M_2 = 3.14$. The S-gals are relatively well balanced, with $M_1/M_2 = 1.32$.

## 7. Structure and Dynamics

We have seen that the formation of our simulated cluster proceeds in a disorderly fashion, with major mergers occurring even at rather recent epochs. At the present time, the cluster remains dynamically active; new clumps continue to fall into its outer parts and even the central regions appear far from equilibrium. We now quantify the dynamical state of the cluster by considering the relative distributions of its various components and the extent to which virial equilibrium is satisfied. We then address the important issue of the accuracy of virial mass estimates.

### 7.1. Spatial Distributions of Dark Matter, Hot Gas and Galaxies

It should be apparent from our earlier discussion that the final spatial distributions of the various constituents of the cluster — diffuse dark matter, hot gas, dark matter halos, G-gals and S-gals — differ in varying degrees. This impression is quantified in Figure 16 which shows the cumulative mass or number density profiles of the various components. The center of the cluster is taken to be at the position of minimum binding energy and the profiles are plotted in terms of the radius normalized to our fiducial cluster value $r_{cl} = 2$ Mpc. Note that the "virial radius", conventionally defined by the sphere which encompasses a mean overdensity of 180, is $r_{vir} = 1.67$ Mpc.

The dark matter profiles from both the gas and star runs are shown in the top panel of Figure 16; the former is slightly more concentrated due to the influence of the central, dominant G-gal. The profiles bend continuously from the center outwards. The spherically averaged, differential density profile of the dark matter from the star run is shown in Figure 17. It is well fit by the function

$$\frac{\rho_{DM}}{\overline{\rho}} = \frac{1500 \; r_{vir}^3}{r \; (5r + r_{vir})^2}, \tag{4}$$

where $\overline{\rho}$ is the mean density of the universe and $r_{vir}$ is the virial radius defined above. This is the same function that fits the scaled dark matter profiles of the clusters in the simulations of Navarro, Frenk & White (1995), but the present simulation has over 20 times more particles and thus resolves the central regions considerably better. To our resolution limit, there is no evidence for a central core radius. Instead, the density continues to increase towards the center, roughly as $\rho_{DM} \propto r^{-1}$ (corresponding to $M(r) \propto r^2$). In the range $0.1 \lesssim r/r_{vir} \lesssim 0.4$, it flattens to an approximate isothermal form, $\rho_{DM} \propto r^{-2}$ and, beyond that, it falls off more steeply, approximately as $\rho_{DM} \propto r^{-2.4}$. In the inner parts, this behavior is similar to that found by Dubinski & Carlberg (1991).

The hot, intracluster gas follows a nearly power-law profile, with slope $-2.2$, over the entire range plotted, $\sim 0.04$ Mpc to 2 Mpc. The profile actually steepens in the center, and the resulting X–ray surface brightness profile resembles that of a cluster with a strong cooling flow. The hot phase is more extended than the dark matter, with a half mass radius (defined within $r_{cl}$) nearly a factor 2 larger. The more extended distribution of the hot gas reflects transfer of energy from the dark matter (Navarro & White 1994; Pearce *et al.* 1994) and transfer of material to the cold phase. The total baryon fraction is dominated by hot gas in the outer parts of the cluster and by cold, G-gal material in the inner few hundred kiloparsecs. Within 2 Mpc of the cluster center, the mean baryon fraction in the gas run is 0.075, *i.e.* 25% smaller than the global value.

The cumulative number distributions of dark matter halos, G- and S-gals in two different mass ranges are shown in the bottom two panels of Figure 16. The dark matter halos are the most extended component, primarily because of the extent of the massive, central halo; in essence, the cluster itself. The S-gals ($N \geq 32$) are more concentrated than the dark mass in the inner regions but, beyond $\simeq 100$ kpc, they closely trace the dark matter. The G-gals are highly concentrated, with a half mass radius nearly a factor three smaller than that of the dark matter. Galaxies of both types exhibit mass segregation, with more massive objects being more centrally condensed. This is a weaker effect for the S-gals. In the body of the cluster, between $\sim 200 - 700$ kpc, the shape of the distribution of S-gals depends only weakly on mass and is approximately unbiased relative to the distribution of dark matter. Beyond this region, the abundance of massive S-gals rapidly drops. Outside the inner 150 kpc, the slope of the S-gal number density profile is $-2.3$, slightly shallower than the profiles typically observed for galaxies in real clusters (*e.g.*, Schombert 1988).

Although the spatial segregation present in the S-gal and G-gal populations is largely a manifestation of dynamical friction (see Section 5) viscous interactions also contribute to mass segregation among the gaseous objects. Such viscous effects are unlikely to be important for real cluster galaxies since their stellar populations typically formed well before cluster collapse. Hence it is encouraging that it is the S-gals rather than the G-gals which are a reasonable match to real clusters. Many real clusters show some evidence for luminosity segregation in their central regions (Capelato *et al.* 1980, Kent & Gunn 1982, Biviano *et al.* 1992, Sodre *et al.* 1992, den Hartog & Katgert 1995) and new spectrophotometric surveys are beginning to uncover evidence for luminosity segregation at larger radii (den Hartog & Katgert 1995). In our simulated cluster mass segregation among the S-gals is mostly confined to the central regions; beyond a few hundred kiloparsecs their distribution is only weakly dependent on mass.

## 7.2. The Question of Hydrostatic Equilibrium

Is the galaxy population close to hydrostatic equilibrium? A simple way to address this question is to examine their distribution in phase space. Figure 18 shows the radial and tangential velocities of the G-gals, S-gals, dark matter halos and a subset of randomly chosen dark matter particles as a function of distance from the cluster center. The figure for the dark matter particles has the phase wrapped appearance characteristic of systems formed via hierarchical clustering, with the (approximate) caustic surface of the most recently accreted mass separating the non–linear portion of the cluster from the outer, quasi–linear infall regime at some radius $r_{nl}$ (Rivolo & Yahil 1983; Bertschinger 1985). The phase-space diagram for the dark matter indicates that the value of $r_{nl} \simeq 3$ Mpc is somewhat larger than our adopted cluster radius of 2 Mpc. The mean interior density contrast at $r_{nl}$ is $\sim 40$.

The dark halos within $r_{nl}$ are predominantly receding from the cluster center. The lack of halos falling in for the second time is due to tidal disruption during the first pericentric passage. In contrast, the G-gals within $r_{nl}$ are nearly all moving radially inward. This is another manifestation of the viscous effects discussed earlier. Galaxies are effectively trapped as they fall through the cluster center, and are prevented from moving back out to large radii. The S-gals are the only population which appears to have little or no net inflow or outflow within $r_{nl}$. Their distribution is similar to that of the dark matter, and they appear to be close to hydrostatic equilibrium.

### 7.3.   Virial Mass Estimates

Dynamical estimates of the masses of rich galaxy clusters, based on the application of the virial theorem or of the equations of stellar hydrodynamics, yield mass-to-light ratios which are typically a factor of 3 to 5 smaller than the ratio of the closure density to the (appropriately weighted) observed mean luminosity density in galaxies (Geller 1984; The & White 1986; Merritt 1987). Proponents of a universe with closure density have long argued that this discrepancy may reflect a bias in the distribution of galaxies towards rich clusters, an idea which was developed formally in the "high peak" model for biased galaxy formation (Davis $et~al.$ 1985, Bardeen $et~al.$ 1986). Partial support for this view was provided by the collisionless simulations of White $et~al.$ (1987) and Frenk $et~al.$ (1988) which showed explicitly how massive galactic halos in the standard CDM model form preferentially in protocluster regions. Other workers (Barnes 1985; Evrard 1987; West & Richstone 1988) pointed out that a further bias could result if dynamical friction segregated galaxies from mass within a cluster. Section 5 showed that all these biases are indeed present in our simulated cluster. We now explore whether they are strong enough to reconcile virial estimates of cluster masses with the theoretical prejudice in favour of $\Omega = 1$.

For real clusters, masses are commonly estimated using the virial theorem which may be written in the form (Heisler & Tremaine 1985):

$$M_{VT} ~=~ \frac{3\pi N}{2G} ~ \frac{\sum_i v_{p,i}^2}{\sum_{i<j} R_{ij}^{-1}}, \tag{5}$$

where $v_{p,i}$ is the line of sight velocity and $R_{ij}$ is the projected separation of a pair of galaxies. Tables 2 and 3 give the mass estimates obtained from this formula using the kinematic data for G- and S-gals within projected radii of 1, 2 and 3 Mpc. Results are given for objects in two separate mass ranges. The quantities $\sigma_{gal}$ and $\sigma_{DM}$ are one-dimensional velocity dispersions, obtained by averaging over three orthogonal projections. The masses quoted are also the mean of three projections, and are given in terms of the true total cluster mass within a sphere of the given radius.

All the virial mass estimates in Tables 2 and 3 underestimate the true mass by an amount which depends on the type of object considered. The heavy G-gals give the smallest virial mass ($\sim 25\%$ of the true value) and the lighter S-gals give the largest ($\sim 75\%$ of the true value). These underestimates result from a combination of two effects. First, all populations of galaxies are "cooler" than the dark matter. As shown in the sixth column of the tables, the magnitude of this "velocity bias" is similar for all populations, ranging from $\sim 30\%$ for the heavier S-gals to $\sim 20\%$ for the lighter S-gals. The second contribution comes from the spatial distributions of the tracer populations which differ markedly among the different types of object. As discussed in Section 7.1, all populations, except the lighter S-gals, are more centrally concentrated than the mass. The ratio of their half-mass radius to the half-mass radius of the dark matter ranges from 0.25 for the heavier G-gals to $\sim 1$ for the lighter S-gals. Since the virial mass estimate is proportional to $R\sigma_{gal}^2$, the velocity bias contributes $20 - 60\%$ of the factor by which the true mass is underestimated and the spatial bias contributes the rest.

When applied to the galaxies in our simulation, the standard "cluster M/L" argument underestimates the mean cosmological density, $\Omega$, by a factor which is at least as large as the factor by which the true cluster mass is underestimated. This is because the mass fraction in galaxies is also biased high relative to the dark matter in the cluster (see §6.3). For example, the mass fractions in lighter and heavier S-gals are biased by factors of 1.5 and 2 respectively, and if these are assumed representative of the global values, these populations yield $\Omega$-estimates of 0.5 and 0.12 respectively. Unfortunately, we cannot determine the global bias in the cluster "light" from our simulation. The region we have modeled is dominated by the cluster and the formation of clumps is suppressed near its apodized edges. These two factors bias our estimates in opposite directions, and the extent to which they cancel out cannot be determined from our calculation alone. Still, our S-gal results suggest that the 'M/L argument" applied to clusters in an $\Omega = 1$ universe may yield misleadingly low estimates of $\Omega$. For the G-gal population and for the galaxy tracers used in earlier work (Carlberg & Couchman 1989, Katz *et al.* 1992; Evrard *et al.* 1994), a similar bias could be blamed on unrealistically strong viscous interactions, but these have no effect on the distribution of our S-gals.

## 8. Discussion

The differences of behavior which we find among our three simulations are a graphic illustration of how relatively small changes in the numerical approach adopted to study galaxy and cluster formation can lead to large quantitative and qualitative changes in the results. The failure of our original experiment to produce a significant number of galaxies demonstrates that, near the dynamic range limit of a gas dynamics code, lack of resolution can significantly reduce the ability of gas clumps to cool and so to make "galaxies"; a relatively modest enhancement of the cooling rate increased the number of dense clumps at the final time by two orders of magnitude. While the amount of cold dense material in "galaxies" appears much more plausible in this second gas simulation, it is possible that resolution effects are still causing a major underestimate of the amount of cold dense gas. Evrard *et al.* (1994) came to a similar conclusion when comparing the cold gas fraction in their own experiment with the much smaller values found in simulations carried out with grid-based hydrodynamics techniques and lower resolution SPH models. Analytic models for galaxy and cluster formation are not, of course, subject to such resolution limitations, and it has long seemed clear that radiative or hydrodynamical heating of pregalactic gas is required to prevent almost all of it from cooling, and so to explain the large amount of diffuse gas observed in galaxy clusters (White & Rees 1978; Cole 1991; White & Frenk 1991; Blanchard, Valls–Gabaud & Mamon 1992; Kauffmann *et al.* 1993; Cole *et al.* 1994).

Once gas is able to cool and to settle into dense clumps, then all experiments agree that at most a small fraction of it is ever reheated or dispersed into the dilute phase by the disruptive effects of later gravitational and hydrodynamical evolution. The high densities attained (which are similar to the observed densities of real galaxies) are sufficient to protect the clumps against tidal or ram pressure stripping. This confirms both the original conjecture of White & Rees (1978) and the earlier numerical results of Carlberg (1991), Hernquist *et al.* (1992), Evrard *et al.* (1994) and Katz *et al.* (1992). Dissipative effects do indeed solve the overmerging problem and allow galaxies to survive the consolidation of their dark halos into a single monolithic cluster halo.

Unfortunately, while all our simulations agree that galaxy disruption is relatively unimportant, they give very different predictions for the strength of frictional drag on galaxy orbits and for the amount of galaxy merging. The latter processes are both much stronger in a simulation where galaxies remain gaseous than in one where they are turned into stars. Viscous drag affects gaseous "galaxies" only in the central regions of the cluster, and their tendency to merge after a close encounter or collision is much greater than that for stellar galaxies of the same size and mass. These effects cause the galaxies in our "gas" simulation to merge rapidly into the central dominant object, and they lead to a final galaxy population which has little resemblance to that seen in most real clusters. It remains possible that, for the minority of clusters which do harbour dominant galaxies containing a large fraction of the total cluster light, mergers of gas–rich progenitors may have played an important role.

By comparison, the evolution of the galaxy population after $z = 0.7$ in our "star" simulation is much milder, and the final system appears a much better model for real systems. It is nevertheless important to remember that at $z > 0.7$ the galaxies are 100% gaseous in this simulation also; their mass function and their spatial distribution may therefore already have been significantly affected by the artificial effects we have just discussed.

It is also possible that the final state of the "star" simulation is determined primarily by dynamical effects occurring after $z = 0.7$ (since this period encompasses the main collapse of the cluster) and that these effects are treated relatively accurately in the model. The galaxies in this model appear to be approximately in equilibrium in the central regions by $z = 0$, even though the crossing time in the cluster is comparable to its dynamical age. Their density distribution is more centrally concentrated than that of the dark matter and, beyond the inner 150 kpc, the profile of the lighter galaxies is quite similar to the profiles observed in real clusters. The stellar "galaxies" exhibit a mild velocity bias and a small amount of segregation by mass. Both these properties are primarily the outcome of dynamical friction. Thus, our simulations lead us to expect some degree of luminosity segregation in real clusters. This is a difficult to measure, but large spectrophotometric surveys are beginning to show convincing evidence for it (Sodre *et al.* 1992, Biviano *et al.* 1992, den Hartog and Katgert 1995).

Applying the virial theorem to the "stellar galaxies" in our simulation leads to an underestimate of the cluster mass which can be quite significant if only the most massive galaxies are used. A further bias is present because more galaxies form per unit mass in the cluster than outside it. Although the strength of this effect cannot be measured accurately in our simulations, it contradicts the common assumption that the mass-to-light ratio of clusters can be identified with the universal value. Thus, a standard $M/L$ analysis of our simulation returns an estimate of $\Omega$ in the range 0.1 to 0.5, even though the true value is unity.

In many respects our simulated cluster appears a plausible match to real clusters. Its bulk X-ray properties are quite similar to those observed, confirming and extending previous results from models in which the IGM was treated as a nonradiative gas. The inclusion of cooling and an admittedly oversimplified prescription for turning cold gas into stars produces a population of galaxies which has many similarities to observed populations. Although the treatment of galaxy formation still needs much improvement, these successes are encouraging and suggest that a viable model for the formation of galaxy clusters is indeed attainable within hierarchical clustering theories such as the one explored in this paper.

This work was supported by a NATO Collaborative Research Grant and by PPARC (CSF and SDMW), NASA grant NAGW–2367 (AEE) and grant of supercomputer time at the San Diego Supercomputer Center sponsored by NSF. The hospitality of the Aspen Center for Physics, where this paper was begun, and the Institute for Theoretical Physics at UC, Santa Barbara, where this paper was finished, is gratefully acknowledged. This research was supported in part by the National Science Foundation under Grant No. PHY89-04035.

**Table 1 :**
**Largest Cluster Members**

| Type | $M_1$ | $M_2$ | $M_3$ |
|---|---|---|---|
| DM halos | 14988 | 440 | 223 |
| G–gals | 5062 | 1612 | 664 |
| S–gals | 728 | 552 | 500 |

**Table 2 : Cluster mass estimates using G–gals**

| $N_p$ | $r_{cut}$ (Mpc) | $N_{gal}$ | $\sigma_{gal}$ (km s$^{-1}$) | $r_h$ (kpc) | $\sigma_{gal}/\sigma_{DM}$ | $M_{VT}/M_{true}(<r_{cut})$ |
|---|---|---|---|---|---|---|
| | 1.0 | 23 | 447 | 152 | 0.82 | 0.41 |
| 32 | 2.0 | 31 | 457 | 240 | 0.84 | 0.44 |
| | 3.0 | 42 | 451 | 392 | 0.83 | 0.56 |
| | 1.0 | 9 | 451 | 95 | 0.83 | 0.27 |
| 128 | 2.0 | 10 | 469 | 117 | 0.86 | 0.23 |
| | 3.0 | 11 | 465 | 141 | 0.86 | 0.22 |

**Table 3 : Cluster mass estimates using S–gals**

| $N_p$ | $r_{cut}$ (Mpc) | $N_{gal}$ | $\sigma_{gal}$ (km s$^{-1}$) | $r_h$ (kpc) | $\sigma_{gal}/\sigma_{DM}$ | $M_{VT}/M_{true}(<r_{cut})$ |
|---|---|---|---|---|---|---|
| | 1.0 | 35 | 467 | 275 | 0.86 | 0.72 |
| 32 | 2.0 | 59 | 450 | 474 | 0.83 | 0.76 |
| | 3.0 | 77 | 431 | 661 | 0.79 | 0.79 |
| | 1.0 | 13 | 371 | 161 | 0.68 | 0.28 |
| 128 | 2.0 | 15 | 374 | 203 | 0.69 | 0.24 |
| | 3.0 | 16 | 368 | 228 | 0.68 | 0.20 |

**Figure Captions**

**Figure 1.** The distribution of a random subset of dark matter particles (left) and gas particles (right) at redshifts $z = 2$ (top), $z = 0.7$ (middle), and $z = 0$ (bottom) located in a comoving square slice half the size of the simulation volume. The slice thickness is 2.25 (physical) Mpc.

**Figure 2.** Evolution of the cluster components: ($a$) dark matter; ($b$) hot gas; ($c$) cold gas; ($d$) G–gals; ($e$) S–gals; and (f) dark matter halos. The different panels show the position of those particles which, at $z = 0$, are contained within a sphere of radius 2 Mpc centered on the cluster. Redshifts are labelled on each panel. A random subset of particles is shown for each component. The definition and properties of the dark matter halos are discussed in Section 6.

**Figure 3.** Grey scale maps of, from left to right, the projected dark matter density, projected baryon density, emission weighted temperature and ROSAT X-ray surface brightness. The maps are generated by placing the cluster at an effective redshift of 0.03 (180 Mpc) from the hypothetical observer. The angular size of the region is $64'$ and the angular resolution is a minumum of $0.5'$. From top to bottom, the epochs shown correspond to redshifts $z = 0.7, 0.3.0.1$ and 0.03, respectively. The spacing between light or dark bands is approximately a factor of two except for the temperature maps where the spacing is $\sim 25\%$.

**Figure 4.** Orbits of ($a$) G–gals and ($b$) S–gals originally identified at $z = 0.7$ which end up in the cluster at the final epoch. The dots mark the positions, in comoving coordinates, of the centre of mass of the particles belonging to a particular object at $z = 0.7$ at intervals of $10^8$ yr. The cross marks the position of the centre of the cluster at $z = 0$. The sixteen most massive objects at $z = 0.7$ are shown, with the mass rank in brackets. Note the extreme differences in orbits for objects in the bottom two rows.

**Figure 5.** Comparison of the orbital properties of G– and S–gals shown in Figure 4. The panels show the distance between the selected object and the position of the most massive G–gal (which tracks the position of maximum baryon density in the gas run), as a function of time. G–gals and S–gals track each other fairly well until their first pericentric passage.

**Figure 6.** Merger histories of the most massive G–gal (left) and S–gal (right) identified at the final epoch. Merging is more extreme in the gas dynamic treatment.

**Figure 7.** "Optical" appearance of the final cluster G–gals (left) and S–gals (right) . A circle is plotted at the projected position of each galaxy, with radius proportional to the square root of its mass. The same scaling is used for both panels, with the smallest circles representing objects of 32 particles.

**Figure 8.** Cumulative binding energy distributions for dark matter particles, S–gals and G–gals at redshifts (from right to left) $z = 0.7, 0.3, 0.1$ and 0. Only material found to be within 2 Mpc of the cluster center at the final time is used.

**Figure 9.** Binding energies of a randomly selected subset of cluster dark matter particles at $z = 0$ and 0.3 compared to $z = 0.7$. The isodensity contours tracing the distribution are used in Figure 10.

**Figure 10.** Binding energies of the particles in the cluster G–gals and S–gals at $z = 0$ and 0.3

compared to $z = 0.7$. All the particles plotted belonged to a "galaxy" at $z = 0$ and at the two redshifts shown. The contours represent the dark matter distribution from Figure 9.

**Figure 11.** (*a*) Number of objects in the simulated volume as a function of redshift for different mass cuts. The solid lines are for G-gals, dotted lines are for S-gals and dashed lines for dark matter halos. (*b*) Number of objects identified using only particles which end up within 2 Mpc of the cluster center at the final epoch. Line types are as in (*a*).

**Figure 12.** Cumulative mass fractions in dark halos, G-gals and S-gals. Each panel shows objects of a given type, with the line styles indicating different mass ranges: $N_p \geq 32$ (solid); $N_p \geq 128$ (dashed) and $N_p \geq 512$ (dotted). (*a*) Entire volume. (*b*) Cluster material.

**Figure 13.** Differential mass functions of galaxies in the entire volume. The best fitting Schechter functions are shown for the G-gals (solid) and S-gals (dashed).

**Figure 14.** Measures of the bias in the G– and S–gals : (*a*) number weighted and (*b*) mass fraction weighted. Line styles are for different mass ranges as in Figure 12. The initial bias in the G–gals is largely lost during the violent mergers which occur after $z = 0.3$. The S–gals preserve their bias.

**Figure 15.** Cumulative multiplicity functions of objects in the cluster (solid lines) compared to predictions based on data from the entire volume. The S–gals are the only population which is overrepresented ('positively biased') within the cluster in both number and mass weighted senses.

**Figure 16.** Normalized cumulative profiles in the final cluster. The top panel shows the dark matter mass profile in the star run (bold) and gas run (light solid) as well as the enclosed total baryon (dotted) and hot gas (dot–dashed) mass profiles. The middle and lower panels show the normalized enclosed number profiles, for two mass ranges, for the G–gals (solid), S–gals (dotted) and dark matter halos (dashed) compared with the enclosed mass profile of the dark matter from the star run.

**Figure 17.** Differential density profile of the dark matter in the cluster. The solid line shows the profile in the star run and the dashed line shows the fitting formula from Navarro, Frenk & White (1994), equation (4).

**Figure 18.** Phase space diagrams for (a subset of) dark matter particles, dark matter halos, G–gals and S–gals. The radial (left) and tangential (right) components of the velocity are plotted against radius from the cluster center at the final time. Circles represent massive objects with $N_p \geq 128$ while dots represent objects with $32 \leq N_p < 128$. Most dark matter halos in the non–linear regime ($r \lesssim 3$ Mpc) are outflowing, while the G–gals in this regime are typically flowing inward. S–gals are the only population with an approximately hydrostatic signature.